\documentstyle[twocolumn,aps,prl,epsf]{revtex}
\begin{document}
\wideabs{

\title{Nanometer Scale Dielectric Fluctuations at the Glass Transition}
\author{E. Vidal Russell, N. E. Israeloff, L. E. Walther and H. Alvarez
Gomariz}
\address{Department of Physics and Center for Interdisciplinary Research
on Complex Systems\\Northeastern University, Boston, MA 02115, USA}
\maketitle

\begin{abstract}
Using non--contact scanning probe microscopy (SPM) techniques, dielectric
properties were studied on 50 nanometer length scales in
poly--vinyl--acetate (PVAc) films in the vicinity of the glass transition.
Low frequency ($1/f$) noise observed in the measurements, was shown to
arise from thermal fluctuations of the electric polarization. Anomalous
variations observed in the noise spectrum provide direct evidence for
cooperative nano--regions with heterogeneous kinetics. The cooperative
length scale was determined. Heterogeneity was long--lived only well
below the glass transition for faster than average processes. 
\end{abstract}

} 

\narrowtext

Glasses serve as the prototype for the kind of slow, non--exponential
relaxation found in diverse systems from magnets and superconductors to
proteins and granular assemblies. Although much has been learned about
glassy phenomena \cite{reviews}, a detailed picture of the underlying
dynamical processes remains elusive. Cooperative dynamics on nanometer
scales has long been postulated \cite{ag,ep}, but never been directly
observed.  A key question, with broad implications, is whether
non--exponential dynamics arises primarily from a heterogeneous
collection of independent nano--scale exponential processes \cite{ag,ep},
or more complex local dynamics \cite{ld}.  Recently, molecular dynamics
simulations revealed the presence near the glass transition of long--lived
clusters surrounded by liquid \cite{clusters}. These mesoscopic scale
clusters exhibited a power--law size distribution. Some macroscopic
experiments have inferred that very long--lived nanoscale structural
\cite{moy} and dynamical \cite{lldh} heterogeneities develop near the
glass transition, while others find very short lived dynamical
heterogeneity \cite{sldh}. Such heterogeneities and their evolution may
play a key role in dynamical scaling behavior \cite{scal}, and possible
dynamical \cite{dt} or phase \cite{ft} transitions, and have implications
for protein dynamics \cite{pd}. 

Since the invention of the atomic force microscope in 1986 \cite{afm}, a
number of related, powerful, scanning probe microscopy (SPM) techniques
have been developed. By sensing local electrostatic forces, non--contact
SPM techniques can be used to measure variations in local dielectric
constants \cite{ldc}.  In this paper, we describe investigations of
dynamics on 50 nm length scales in poly--vinyl--acetate (PVAc) films near
the glass transition, $T_G \sim 306$K, via measurements of dielectric
properties using an ultra--high--vacuum, variable temperature SPM.  We
observed low--frequency `$1/f$' fluctuations, which are shown to arise
from thermal polarization noise, and can be used to study the equilibrium
nano--scale dynamics.  Anomalous temporal variations in the spectral
dependence of the noise were found. The variations were used to determine
cooperative length scales and related to the time evolution of dynamical
heterogeneity.

In probing a mesoscopic volume of a structural glass, deviations from
macroscopic response might be observed in, for example, the
frequency--dependent dielectric susceptibility, $\varepsilon (\omega ) =
\varepsilon'(\omega ) + i\varepsilon ''(\omega )$ or similar response or
correlation (e.g. noise) functions. In the simplest picture \cite{ag},
independent nano--regions of cooperativity of typical size $\xi$
\cite{reviews,ag,ep,clusters,moy,lldh} relax exponentially (Debye--like
response), but have a distribution of characteristic relaxation times,
$\tau$. Estimates for $\xi$ based on measurements \cite{lldh} and theory
\cite{ag} have been in the range 2 -- 5 nm for various glass--formers at
$T_G$. Far from the $\alpha$--peak in $\varepsilon ''$, only those
nano--regions with $\tau \sim 1/\omega$, will contribute significantly to
$\varepsilon ''(\omega ,T)$. By using the Kramers--Kronig relations, it
is easily shown that $\xi$ relates to the number of nano--regions,
$N(\omega)$, which contribute near $\omega$ within a band,
$\Delta\omega\sim\omega$ in width (i.e. factor of e) in a sample of volume
$\Omega$, via: 

\begin{equation}
\xi^3 \approx {\Omega\varepsilon ''(\omega ,T) \over N(\omega)\varepsilon
' (0, T)}
\end{equation}

\noindent 
When $N(\omega )$ is of order 10 \cite{dev}, deviations from bulk--like
behavior will be clearly observable, for example, anomalous variations in
$\varepsilon ''$ (or equivalent measured quantity) as a function of
frequency. These spectral anomalies would be {\sl persistent} for very
long--lived dynamical heterogeneity. Thus a lower--limit on the sample
volume needed is of order: $\Omega\sim 10\xi^3\varepsilon ' (0,T)/
\varepsilon ''(\omega ,T)$. For $\varepsilon ''\sim 0.1$ below $T_G$, we
estimate $\Omega\sim 1\times 10^{-16}$ cm$^3$ for PVAc, e.g. a cube of
about 50 nm on a side. If nano--regions are not independent \cite{ld} or
undergo intermittent or evolving dynamics \cite{pd,dev}, or if the
amplitude of the response varies significantly from region to region,
mesoscopic effects might be observed in larger samples. 

Complex dynamics have been studied in great detail using noise
spectroscopy in mesoscopic {\sl conductors} at low temperature
\cite{dev,mc}. A sensitive mesoscopic signature involves anomalous
statistical variations in the noise power, beyond the expectations of
Gaussian statistics, i.e. the noise is non--Gaussian \cite{dev}. It was
recently demonstrated that $1/f$ noise can be measured in dielectric
materials near the glass transition \cite{dmgt}. This noise arises from
thermal polarization fluctuations, which relate to $\varepsilon (\omega)$
via the fluctuation--dissipation theorem (FDT). However, extending noise
measurements to the mesoscopic scale in dielectrics, as was done in
conductors, is not possible with these techniques \cite{dmgt}, due to 
overwhelming technical limitations. 

In the present experiments, we employed a non--contact capacitance
measurement scheme \cite{ldc} using a thermal drift compensated,
ultra--high--vacuum, variable temperature SPM \cite{aptech}. A small
piezo--resistive cantilever \cite{cant} with a sharp conductive tip was
driven near its resonance frequency in high vacuum close to the 0.5 $\mu$m
thick sample films, which were spin--coated onto a metal substrate. 
Details of the instrument design and PVAc film preparation are discussed
elsewhere \cite{aptech}. When a voltage bias is applied between tip and
the substrate (see Fig. 1 inset), the resonance frequency, $f$, of the
cantilever will decrease due to the electrostatic forces \cite{ldc}. At
fixed height, variations in the tip--substrate capacitance or dielectric
constant can be directly related to variations in the resonance frequency. 
Since the second derivative of the capacitance is responsible for the
resonance frequency shifts, stray capacitances play a negligible role for
tip heights $< 100$ nm.

\begin{figure}
\leavevmode
    \epsfxsize=3.3in\epsfbox{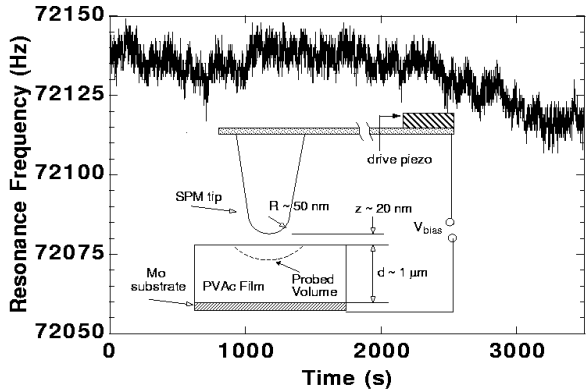}
\vskip0.3cm 
\caption{Resonance frequency vs. time after equilibration with 4V bias,
tip height 20 nm, at 305K is shown. Inset: measurement configuration.}
\label{fig1}
\end{figure}

Using a hemispherical tip model \cite{ldc} at a distance $z < r$, the tip
radius, we have calculated $C (z)$ and $\Omega$ (effective probed volume).
For example for a typical tip of radius 50 nm, held 20 nm from the
surface, with $\varepsilon =2$, we find $C = 8\times 10^{-18}$ F, and
$\Omega \sim 1.3\times 10^{-16}$ cm$^3$. It is important to note that
subsurface material is probed to a depth ($\sim$ 50 nm) significantly
larger than the length scales below which surface effects would be
expected to dominate the dynamics, as in the experiments on glassy
dynamics in porous media, with pores $< 10$ nm \cite{pm}. 

In earlier studies of PVAc we showed that dielectric relaxation could be
studied using the SPM resonance frequency shift \cite{aptech}. Nearly
bulk--like stretched--exponential relaxation was observed on 50 nm scales,
and could be used to clearly identify the glass transition and
characterize relaxation times, which increased rapidly on cooling through
the glass transition. In the course of these relaxation studies, we
observed a distinct background noise on relaxation curves or in
equilibrium, which appeared only when the sample films were present and
only when the tip was close to the film surface \cite{meson}. See figure
1. PMMA films had an order of magnitude lower noise power than the PVAc
films near room temperature ($T_G \sim 110^{\circ}$C for PMMA). A variety
of checks were carried out to rule out instrumental or other extrinsic
sources of noise. 

\begin{figure}
\leavevmode
   \epsfxsize=3.3in\epsfbox{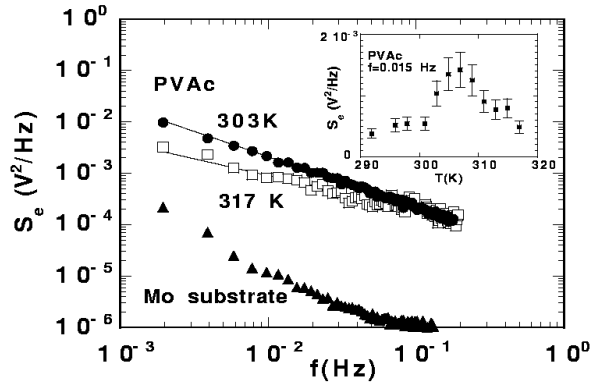}
\vskip0.3cm 
\caption{Voltage noise spectra extracted from SPM resonance frequency
fluctuations via eqn. 2 are shown for $T=302.9$K in PVAc and for bare Mo
substrate. An instrumental rolloff occurs above 0.2 Hz. Inset: temperature
dependence of voltage noise at $f=0.0 15$ Hz vs. temperature for PVAc is
shown.}
\label{fig2}
\end{figure}

In order to understand these fluctuations, recall that the SPM senses
tip--sample capacitance by measuring local electrostatic forces.  The
system can be modeled as a dielectric filled capacitor ($C_D$) in series
with a vacuum capacitor ($C_V$) all in parallel with the much larger
stray capacitance, $C_S$. Thermal fluctuations in the dielectric film
\cite{dmgt} will produce a fluctuating emf with spectral density, $S_e$,
across $C_D$. Voltage noise appearing across the series combination must
sum to near zero, since it is ``shorted out'' by the large parallel $C_S$,
and would not be detectable by any voltmeter. Thus, voltage noise close to
$S_e$ will appear with opposite sign across $C_V$. The tip will therefore
feel fluctuations in electric field and thereby force. This acts
approximately as a fluctuating voltage, with spectral density $\sim S_e$,
added to the constant voltage bias, and the resonance frequency will
fluctuate. The spectral density of the thermal resonance frequency noise
will be approximately: 

\begin{equation}
S_f=\Biggl({\partial f \over \partial V}\Biggr)^2 G(\varepsilon ) S_e =
\Biggl({\partial f \over \partial V}\Biggr)^2 G(\varepsilon) {4 k_{\rm B}
T \varepsilon '' \over |C|^2 \omega}
\end{equation}

\noindent
where $\partial f/\partial V$ is the measured shift in resonance frequency
for small changes in bias voltage at the operating point, $k_{\rm B}$ is
Boltzmann's constant, $\omega$ is the fluctuation frequency, and
$G(\varepsilon)$ is a dimensionless geometrical factor which is of order
unity and depends weakly on dielectric constant \cite{meson}. 

Long time series of the resonance frequency were recorded with fixed
conditions ($V=8$V), using feedback to periodically reset the resonance
frequency (every 500 s). These time series were Fourier analyzed and
averaged to produce a power spectrum. Power--law, $1/f^{\alpha}$,
spectra were observed. By inverting eqn. 2, $S_e$ could be calculated from
the measured noise. See fig. 2.  We used $G(\varepsilon) = 5$ at low
temperature decreasing monotonically to 2.7 at high temperature, estimated
based on a parallel plate model. At temperatures above the glass
transition, the noise power decreased and the spectrum flattened as
expected based on the known behavior of $\varepsilon$ \cite{dmgt}. A peak
in the noise just below the glass transition is predicted, and is observed
(see fig. 2 inset). Using our estimates for capacitance, and bulk values
for $\varepsilon$ \cite{deps}, eqn. 2 predicts a peak voltage noise at
0.015 Hz of $S_e\sim 4\times 10^{-3}$ V$^2$/Hz. The measured $S_e$ peaks
at $1.5\times 10^{-3} $ V$^2$/Hz (inset), excellent agreement given the
approximations involved. In addition, explicit expressions for the noise
in terms of bias voltage and resonance frequency can be derived, and were
found to be consistent with the measurements \cite{meson}. This resonance
frequency noise is unrelated to the thermo--mechanical noise \cite{cant},
which can be observed at the resonance frequency.

\begin{figure}
\leavevmode
   \epsfxsize=3.3in\epsfbox{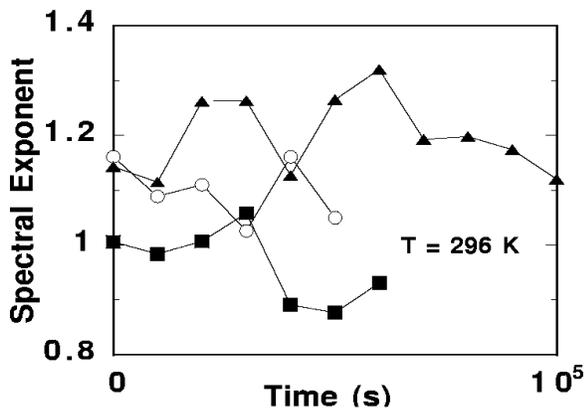}
\vskip0.3cm 
\caption{Noise spectral exponents, measured between $f=0.01$ -- 0.08 Hz,
vs. time for succesive measurements at 296K are shown.}
\label{fig3}
\end{figure}

In order to study variations in the spectral dependence, the local
spectral exponent, $\gamma=-\partial \ln S/\partial \ln f$, was measured
in the 0.01 -- 0.08 Hz band for spectra averaged over 20 FFTs, and was
studied as a function of time at various temperatures. The results of
this analysis were quite revealing, as shown in figure 3. At 296K,
$\gamma$ showed anomalous variations between succesive measurements, i.e.
on a time scale of about 10$^5$ s. Similar but smaller anomalous
variations were observed at 298K. Figure 4 shows variance of the
spectral exponent vs. temperature. Well below and well above this range
of temperature, at 292K and 303K respectively, $\gamma$ exhibited little
variation with time beyond the statistical variations found in Gaussian
noise or in Monte--Carlo simulations of noise produced by random
distributions of two--state exponential processes with fixed
characteristic times, (shown as dashed line) \cite{meson}. Also shown
(inset) is the autocorrelation function for the spectral exponent at 296K.
The characteristic decay time is $6\times 10^4$ s.  By comparison, the
measured relaxation time was $3\times 10^4$ s at 296K \cite{aptech}.
Variations were also studied on a shorter observation time scale of about
$2\times 10^4$ s, at several temperatures. For this shorter time scale,
anomalously large variations occured only at higher temperatures, 298K and
301K \cite{meson}. 

\begin{figure}
\leavevmode
   \epsfxsize=3.3in\epsfbox{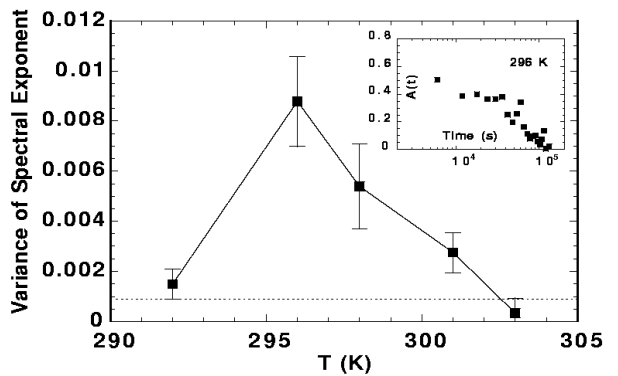}
\vskip0.3cm 
\caption{Variance of spectral exponent (0.01 -- 0.08 Hz) measured in a
time window centered on $1\times 10^5$ s is shown vs. temperature. Also
shown is a variance for simulation of fixed distribution of exponentials
(dotted line). Inset: autocorrelation function for exponent at 296K.}
\label{fig4}
\end{figure}

The variations observed in the noise spectrum suggest a superposition of a
small number of discrete components with slowly evolving kinetic
parameters. Below $T_G$, memory of the local kinetic parameters is lost on
a characteristic time scale about twice the average relaxation time,
consistent with NMR results \cite{sldh} measured well above $T_G$. The
matching of the average relaxation time (at 296K) to the experimental
observation time explains the peak in variance at 296K. This time--scale
rapidly becomes very large on cooling, thus little variance is observed
at 292K. Thus dynamical heterogeneity becomes relevant, a few degrees
below $T_G$, for those fluctuations which occur on time scales much
shorter than the average. We found that we could produce similar anomalous
spectral exponent variations in the simulations of distributions of
exponentials by randomizing the characteristic times \cite{meson}. The
size of the variance matched the 296K value when the density of
exponential processes was, $N(\omega) = 7$. Using eqn. 1, we find $\xi =
10\pm 4$ nm at this temperature, somewhat larger than conventional
estimates \cite{lldh}. 

In summary, we described a new method of probing equilibrium nanoscale
glassy dynamics via electric polarization fluctuations. Anomalous temporal
variations were observed in the noise spectral dependence below the glass
transition in PVAc, a direct indication of cooperative nano--regions with
heterogeneous but evolving kinetics. We determined the cooperative length
scale, and showed that dynamical heterogeneity persists for times
comparable to the average relaxation time, for faster than average
processes. This kinetic evolution is qualitatively consistent with models
in which the dynamics of neighboring nano--regions are highly coupled
\cite{ld}. Extending the measurement bandwidth, using FM techniques
\cite{fm}, will facilitate the use of noise statistics to analyze in
greater detail, the local dynamics for comparison with various models and
simulations. 

We thank S. R. Nagel and D. Braunstein for helpful discussions, and M.
Tortonese and Park Scientific for the piezolevers. Supported by the NSF
Grant no. DMR 9458008.

\end{document}